\documentclass[12pt]{irsreport}
\usepackage{epsfig}

\title{IRS-TR 12001:  Spectral Pointing-Induced Throughput Error
and Spectral Shape in Short-Low Order 1}

\author{
G.C. Sloan (1) ~\& D. Ludovici (2) \thanks{ (1) Infrared Spectrograph 
Science Center, Cornell University, (2) Department of Physics and
Astronomy, University of Iowa; NSF REU Research Assistant, Astronomy 
Department, Cornell University} }

\date{14 December, 2012}

\runninghead{IRS-TR 12001 - SPITE and Shape in SL1}

\begin{document}

\maketitle

\begin{abstract}
We investigate how the shape of a spectrum in the Short-Low
module on the IRS varies with its overall throughput, which
depends on how well centered a source is in the spectroscopic 
slit.  Using flux ratios to quantify the overall slope or
color of the spectrum and plotting them vs.\ the overall
throughput reveals a double-valued function, which arises 
from asymmetries in the point spread function.  We use this
plot as a means of determining which individual spectra are
valid for calibrating the IRS.
\end{abstract}

\section{Introduction} 

The Infrared Spectrograph (IRS; Houck et al.\ 2004) on the 
{\it Spitzer Space Telescope} (Werner et al.\ 2004), like all
slit spectrometers, suffers from the issue of slit 
throughput.  As the position of a source within the slit
varies, the amount of radiation from the source truncated
by the edges of the slit also varies.  The point-spread
function (PSF) increases in size with wavelength, and the
interactions of the Airy rings in the PSF with the slit
edges make the throughput a complicated function of both
telescope pointing and wavelength.  

The IRS Team at Cornell has referred to this problem as 
Spectral Pointing-Induced Throughput Error, or SPITE.  Even 
before the launch of the telescope in August, 2003, they 
were investigating both the impact of SPITE and its possible 
mitigation.  Through the course of the mission, several
IRS Technical Reports have addressed these issues.  The
very first of the reports, released a month ahead of launch,
characterized the likely effect SPITE would have on the
shape of observed spectra in all four IRS modules (IRS-TR
03001, Sloan et al.\ 2003).  IRS-TR 04007 (Keremedjiev \& 
Sloan 2004) found that the reconstructed pointing from the
spacecraft did not have the accuracy needed to directly
determine SPITE corrections, but IRS-TR 04005 (Sloan 2004)
noted that the orthogonality of the Short-Low (SL) and
(LL) slits made it possible to use the position of a
source in the cross-dispersion direction (i.e. along the 
slit) in one module to estimate its position in the
dispersion direction (across the slit) in the other.

IRS-TR 06001 (Keremedjiev \& Sloan 2006) provided perhaps
the most important result, that the distortion to the shape 
of the the SL Order 1 (SL1) spectrum was generally less than 
2\% for most pointings (i.e.\ pointings that caused a total
loss of flux of 8\% or less, compared to pointings in the
center of the slit).  The implication is that scalar
corrections would generally suffice to produce accurate
spectral shapes.

\section{Observations and Analysis} 

\bigskip
\begin{center}
\begin{tabular}{lllr} 
\multicolumn{4}{c}{\bf Table 1---Calibration Data} \\
\hline
{\bf Standard} & {\bf Spectral} & {\bf Observed in} \\
{\bf star} & {\bf class} & {\bf campaigns} & {\bf AORs} \\
\hline
HR~6348      & K1 III & P--61 (intermittently) &   84 \\
HD~166780    & K4 III & P--61 (intermittently) &   41 \\
HD~173511    & K5 III & P--61                  &  132 \\
$\alpha$ Lac & A1 V   & 1--58 (intermittently) &   28 \\
$\delta$ UMi & A1 V   & P--19 (intermittently) &   43 \\
\hline
\end{tabular}
\end{center}
\bigskip

This report concentrates on the large number of pointings
to the five stars listed in Table~1.  These five stars are
the core standards used by the IRS Team at Cornell for the
low-resolution IRS modules.  They account for 327 of the
calibration observations with the IRS.  Among other things,
they provide us with a large sample of pointings randomly
distributed about the center of the spectroscopic slits.

We will focus on the 1st-order SL slit (SL1).  SL is much 
more susceptible to pointing errors than LL because its slit 
is $\sim$3 times narrower, 3$\farcs$6 vs.\ 10$\arcsec$, 
compared to typical pointing errors of $\sim$0$\farcs$5.  
Within SL, the larger PSF in SL1 means that SPITE will be 
more of a problem for typical pointing errors than in SL2.

Our immediate objective is identifying those SL pointings
that should not be included when combining spectra from the
many observations for calibration purposes.  In the process,
we have uncovered evidence that the PSF is asymmetric in the
dispersion direction, with clear consequences for which
spectra have distorted shapes and should not be used in
calibration.

Each Astronomical Observing Request (AOR) includes two 
pointings in SL1, which we treat as separate measurements.  
For each pointing, we measure the mean flux density from 7.5 
to 14.0~\mum\ to assess the total throughput from the star 
(F$_{\nu}$ units, weighted by a 1/$\lambda^2$ to evenly weight 
data at all wavelengths).  To estimate the impact of SPITE on 
the shape of the spectrum, we also consider the ratios of the 
mean flux density in the following three wavelength regions:
7.5--9.5, 9.5--11.5, and 11.5--14.0~\mum.  

Figures~1 and 2 plot the flux ratios for the intervals
11.5--14.0~\mum\ over 7.5--9.5~\mum\ as a function of the
mean flux density over the larger 7.5--14.0~\mum\ window.  
For each source, we normalize the results by ignoring the 
three brightest spectra and averaging the next five.  To 
simplify our discussion, we will refer to the quantities 
plotted as color vs.\ flux.

\begin{figure} 
  \begin{center}
     \epsfig{file=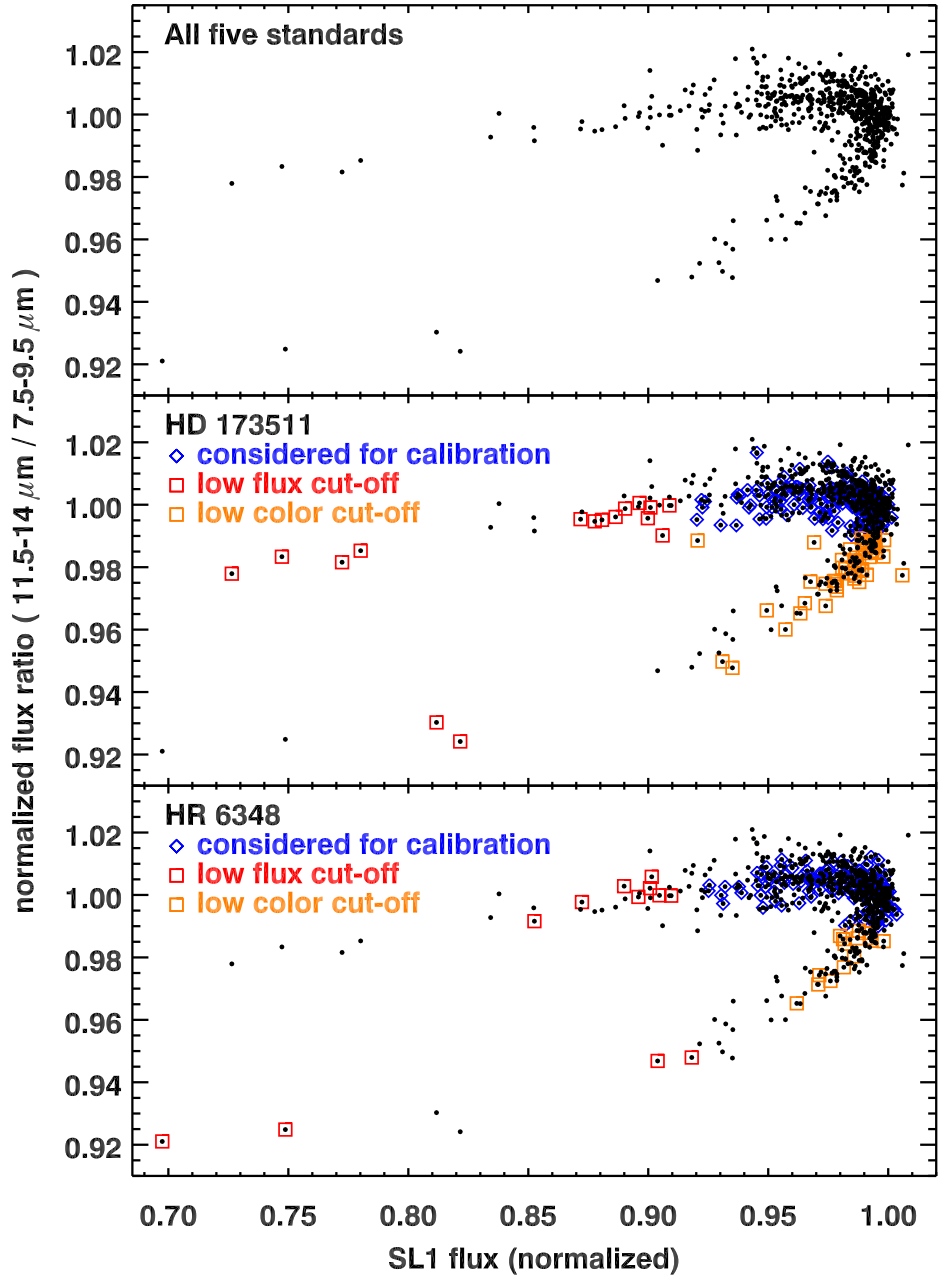, width=12.5cm}
  \end{center}
\caption{
---Normalized color vs.\ flux for (top to bottom):  all five 
standard stars considered, HD 173511, and HR 6348.  The data 
for a particular star are color coded based on whether they 
fail the flux criterion (0.92), the color criterion (0.99),
or pass both.}
\end{figure}

\begin{figure} 
  \begin{center}
     \epsfig{file=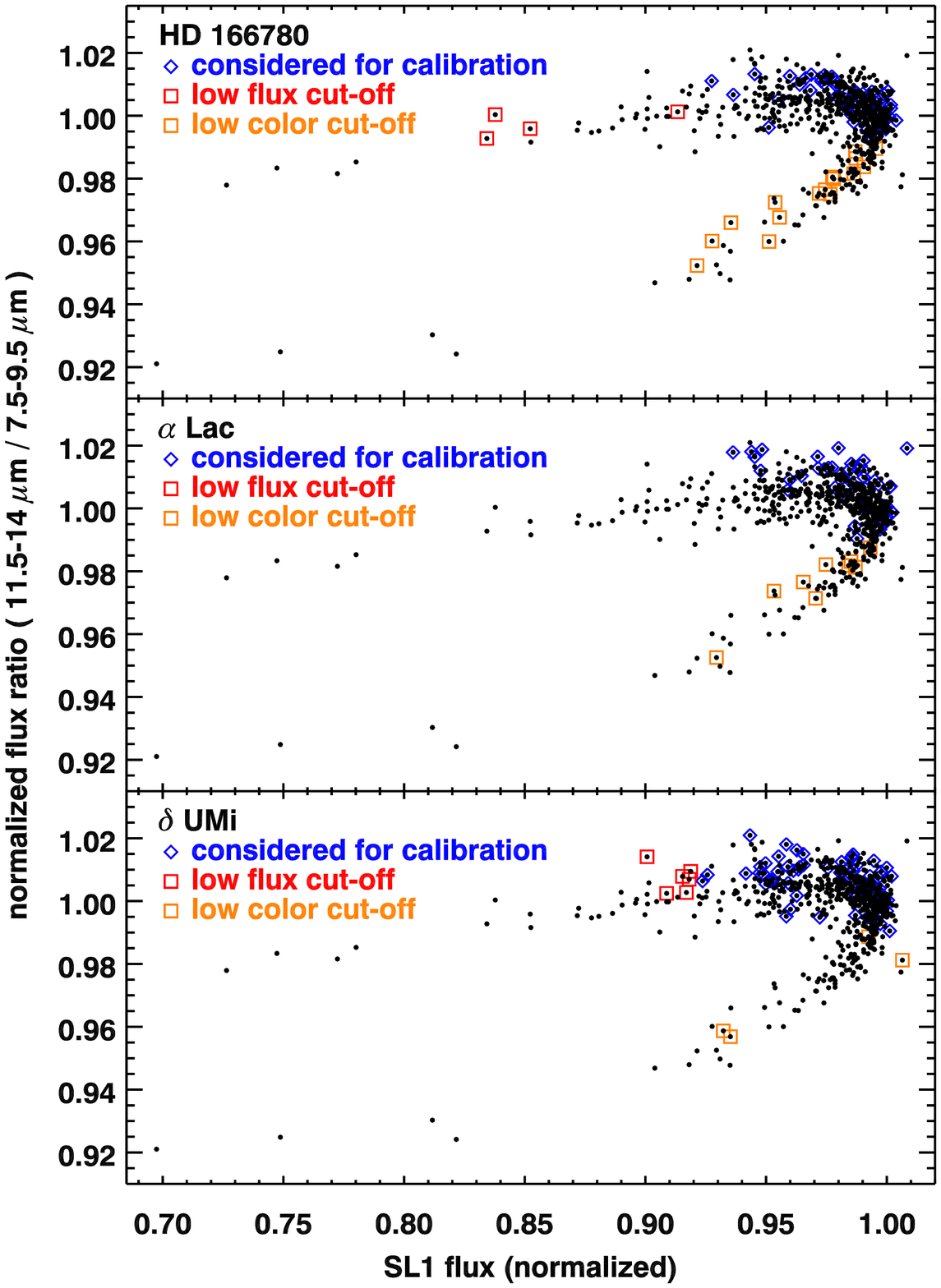, width=12.5cm}
  \end{center}
\caption{
---Normalized color vs.\ flux for (top to bottom):  HD 166780, 
$\alpha$ Lac, and $\delta$ UMi.  The symbols are as defined in
Fig.~1.}
\end{figure}

\section{Discussion} 

Figure 1 and 2 reveal a double-valued dependence of color
with flux, with two significantly different colors possible
for a given reduction in flux.  Following the upper trace,
as the flux drops, the color actually grows redder before
beginning a steady, but slow change to the blue.  Even at
92\% relative throughput, the change in color is less than 1\%.
The lower trace, however, quickly grows bluer as the overall
throughput drops.  

The general trend to bluer colors with less throughput 
results from the increasing size of the PSF with wavelength.
An offset of the source from the center of the slit 
preferentially removes flux from the red end of the spectrum
compared to the blue.  

IRS-TR 03001 predicted the small increase in red flux
compared to the blue for slight offsets from the center of
the slit.  At long enough wavelengths, the first Airy ring
is partially truncated by the slit edges.  Shifting the 
source to slightly off-center pulls the Airy ring on one 
side back into the slit.  If the PSF is asymmetric, then this 
effect may be more pronounced for shifts in one direction 
compared to the other.

To demonstrate that the two arms in Fig.~1 and 2 arise from 
offsets in different directions in the dispersion direction 
in SL, we can take advantage of the nearly orthogonal angles 
between the long axes of the slits in SL and LL.  IRS-TR 
04005 showed that the pointing errors accumulated between the 
SL and LL observations were small enough to allow the 
position {\it along} the LL slit to serve as a proxy for 
position {\it across} the SL slit.  

The worst four pointings on the upper arm in Fig.~1 and 2
correspond to two HD~173511 AORS with a mean offset of 
$-$0.16 pixels from the nominal nod position in LL2, where 
we define zero as the left-hand side of the LL spectral 
images.   The worst four pointings on the lower arm are from
two AORs, one of HD~173511 and one of HR~6348; the mean 
offset here is $+$0.23 pixels.  Translating to angles on the
sky, these are $-$0$\farcs$79 and $+$1$\farcs$17, 
respectively.  

The next worst group of pointings on the two arms give 
similar results.  Here we will use the next six pointings,
using flux as the criterion on the upper arm and color as the
criterion on the lower.  The six pointings on the upper arm 
are from five AORs of HD~166780, HR~6348, and HD~173511, and
their mean offset is $-$0.05 pixels, or $-$0$\farcs$26.  The
six pointings on the lower arm are from four AORs.  Three are
of the same sources just listed; the fourth is of 
$\alpha$~Lac.  Their mean offset is $+$0.06 pixels, or 
$+$0$\farcs$29.  It is worth noting that even this far
away from the center of the distribution, the offsets in LL2
are beginning to overlap.  The implication, unfortunately, is
that the pointing errors accumulated during an AOR will be
large enough to prevent us from predicting the throughput in
SL from the LL positions with enough accuracy to be useful in
all but the most extreme cases.

\bigskip
\begin{center}
\begin{tabular}{llrrrr} 
\multicolumn{6}{c}{\bf Table 3---Rejected and Retained Pointings} \\
\hline
{\bf Standard} & {\bf Spectral} & {\bf SL1} & {\bf Below} & {\bf Below} \\
{\bf star} & {\bf class} & {\bf pointings} & {\bf flux limit} & {\bf color 
  limit} & {\bf Retained} \\
\hline
HR~6348      & K1 III & 168 & 12 &  15 & 151 \\
HD~166780    & K4 III &  82 &  4 &  17 &  61 \\
HD~173511    & K5 III & 264 & 16 &  58 & 190 \\
$\alpha$ Lac & A1 V   &  56 &  0 &   9 &  47 \\
$\delta$ UMi & A1 V   &  86 &  6 &   4 &  76 \\
\hline
\end{tabular}
\end{center}
\bigskip

\begin{figure}[h!] 
  \begin{center}
     \epsfig{file=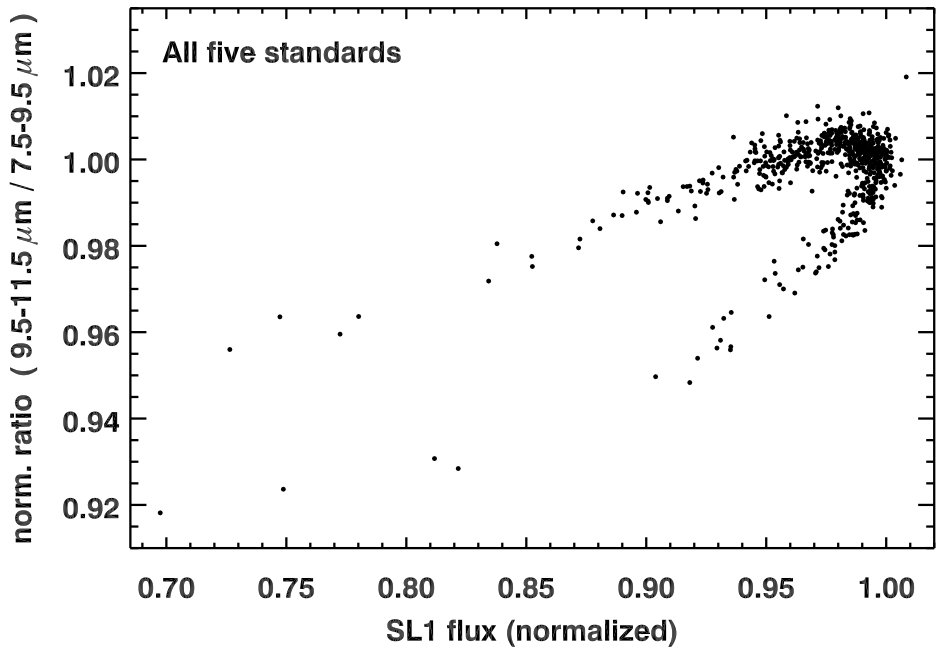, width=12.5cm}
  \end{center}
\caption{
---As Fig.~1 and 2, except that the color intervals have been
shifted to 9.5--11.5~\mum\ (the central interval) and
7.5--9.5~\mum\ (the blue interval).  The flux in the central
interval decreases more than the red or blue intervals,
resulting in a sharper drop in color with decreasing flux 
compared to Fig.~1 and 2.}
\end{figure}

The distribution of fluxes and colors for the sources provide
a quantitative means of consistently identifying mispointed
observations and excluding them from the calibration of the
SL module.  We have chosen a normalized flux limit of 0.92,
based originally on IRS-TR 06001.  In order to maintain good
consistency in the end-to-end slope of the SL1 data, we will
reject any pointings below a normalized color of 0.99.  The 
sources which fall below these limits are color coded in 
Fig.~1 and 2.  Table~2 provides counts of rejected pointings
for each of our five primary standards.  Observations which 
pass this test might still be rejected if visual inspection
reveals other artifacts or problems.

Figure~3 presents a color-flux plot similar to Fig.~1, except 
that the color has been shifted to the ratio of the central 
interval (9.5--11.5~\mum) over the blue interval 
(7.5--9.5~\mum).  This version of the color-flux plot shows
similar bahavior as before, except that this color decreases
more quickly as the overall flux decreases.  The cause is a
greater loss of flux in the central wavelength region in SL1
than at either end of the spectrum.  The {\it Spitzer} 
Science Center refers to this problem as ``Curvature in 
SL1'' in the IRS Instrument Handbook (see Sec.\ 7.3.3).  It 
is only an issue for significantly mispointed spectra.  As 
Fig.~3 illustrates, spectra on the upper arm do not show a 
curvature of more than 1\% unless the overall throughput is 
down by $\sim$10\%.  On the lower arm in Fig.~3, the 
curvature is more of a problem, but that arm is not very well 
populated.  For our selection criteria of 0.92 in overall 
flux and 0.99 in color (11.5--14~\mum/7.5--9.5~\mum), the 
curvature is never worse than $\sim$2\%.


\end{document}